\documentclass[modern]{aastex63}
\usepackage[T1]{fontenc}
\usepackage{multirow,amsmath}

\definecolor{DarkGreen}{rgb}{0.0,0.4,0.0}  % define a custom color

\usepackage[normalem]{ulem}
\usepackage{soul}
\usepackage{cancel}

\shorttitle{Nonparametric Statistics of Erupting Magnetic Flux Ropes}
\shortauthors{Liu \& Wang}

\begin{document}

\title{Nonparametric Statistics on Magnetic Properties at the Footpoints of Erupting Magnetic Flux Ropes}
\correspondingauthor{Rui Liu}
\email{rliu@ustc.edu.cn}

\author[0000-0003-4618-4979]{Rui Liu}
\affiliation{CAS Key Laboratory of Geospace Environment, Department of Geophysics and Planetary Sciences, University of Science and Technology of China, Hefei 230026, China}
\affiliation{CAS Center for Excellence in Comparative Planetology, University of Science and Technology of China, Hefei 230026, China}
\affiliation{Mengcheng National Geophysical Observatory, University of Science and Technology of China, Mengcheng 233500, China}

\author[0000-0002-9865-5245]{Wensi Wang}
\affiliation{CAS Key Laboratory of Geospace Environment, Department of Geophysics and Planetary Sciences, University of Science and Technology of China, Hefei 230026, China}
\affiliation{CAS Center for Excellence in Comparative Planetology, University of Science and Technology of China, Hefei 230026, China}

\begin{abstract}
It is under debate whether the magnetic field in the solar atmosphere carries neutralized electric currents; particularly, whether a magnetic flux rope (MFR), which is considered the core structure of coronal mass ejections, carries neutralized electric currents. Recently Wang et al. (2023, ApJ, 943, 80) studied magnetic flux and electric current measured at the footpoints of 28 eruptive MFRs from 2010 to 2015. Because of the small sample size, no rigorous statistics has been done. Here, we include 9 more events from 2016 to 2023 and perform a series of nonparametric statistical tests at a significance level of 5\%. The tests confirm that there exist no significant differences in magnetic properties between conjugated footpoints of the same MFR, which justifies the method of identifying the MFR footpoints through coronal dimming. The tests demonstrate that there exist no significant differences between MFRs with pre–eruption dimming and those with only post-eruption dimming. However, there is a medium level of association between MFRs carrying substantial net current and those produce pre-eruption dimming, which can be understood by the Lorentz-self force of the current channel. The tests also suggest that in estimating the magnetic twist of MFRs, it is necessary to take into account the spatially inhomogeneous distribution of electric current density and magnetic field.  
\end{abstract}

%\keywords{}

\section{Introduction} \label{sec:intro}
Models for solar eruptions can be roughly categorized into two groups: those enlisting magnetic reconnection as the key process, such as the tether cutting model \cite[]{Moore2001} and the breakout model \cite[]{Antiochos1999}, and those enlisting ideal MHD processes of a current-carrying magnetic flux rope (MFR) as the key process, such as the kink instability \cite[]{Torok&Kliem2005}, the torus instability \cite[]{Kliem&Torok2006}, and the MHD catastrophe \cite[]{Forbes&Priest1995}. These ideal processes would not take effect if there is no net electric current flowing through the MFR. 

Whether coronal magnetic field is current neutral has been controversial \cite[e.g.,][]{Melrose1991,Parker1996}. Recent three-dimensional numerical experiments suggest that current non-neutralization would develop in active regions driven by either flux emergence \cite[]{Torok2014} or photopsheric flows \cite[]{Dalmasse2015}. Meanwhile, observational studies indicate that active regions carrying net current might be more CME-productive than current-neutral ones \cite[e.g.,][]{Liu2017,Avallone&Sun2020,Liu2024}. It is also remarkable that the degree of current non-neutrality is found to be a proxy for assessing CME productivity as good as the magnetic shear at polarity inversion lines \cite[]{Liu2024}. However, it is still unclear whether an MFR, which is generally considered the core structure of CMEs, carries substantial net current. \cite{Liu2016} showed that an MFR identified in the extrapolated pre-eruption coronal field is associated with strong electric currents. On the other hand, \cite{Wang2017} found that an MFR formed during the eruption is anchored in regions with weak electric currents (their Supplementary Figure 3) and carries minimal net current (their Supplementary Figure 4).

Although direct measurements of the coronal magnetic field are unavailable, coronal MFRs are rooted in the dense photosphere, where the vector magnetic field is now routinely measured by the Helioseismic and Magnetic Imager \citep[HMI;][]{Scherrer2012} onboard the Solar Dynamics Observatory \citep[SDO;][]{Pesnell2012}. Furthermore, the footpoints of an eruptive MFR are often outlined by hooked part of flare ribbons and/or manifested by coronal dimmings \cite[]{Qiu2007,Janvier2014,Wang2017,WangW2019,Gou2023}. With the footpoints being identified, the magnetic field and electric current are supposed to `flow' from one footpoint throughout the MFR to the other, therefore providing valuable information on the properties and evolution of the eruptive MFR. \citet[][hereafter Paper I]{Wang2023} carried out a survey of footpoint properties of 28 MFRs, which are all associated with CMEs. The events are selected from the database RibbonDB \cite[]{Kazachenko2017}, which include two-ribbon flares of GOES class C5.0 and larger within $45^\circ$ from the disk center from 2010 April to 2016 April. The selected 28 events are considered to possess an MFR structure because of observational signatures including the classical three-part structure or twisted loop-like structure observed in white-light coronagraphs, magnetic clouds observed in situ by near-Earth spacecrafts, and MFR proxy structures observed prior to eruption in the corona, such as sigmoids, hot channels, filaments, and expanding coronal loops (Table 1 in Paper I). 

\begin{figure}[htbp] \label{fig:example}
	\centering
	\plotone{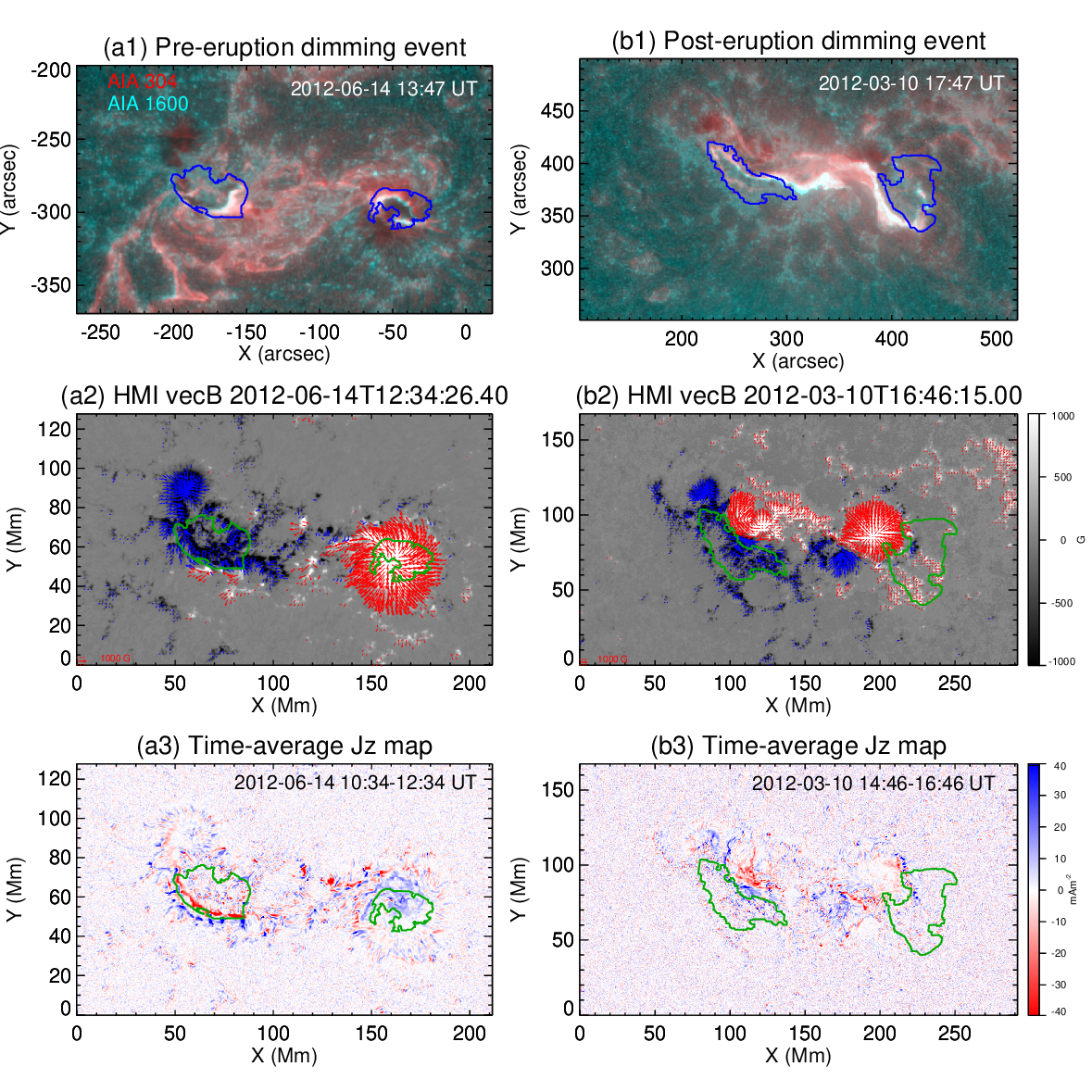}
	\caption{Footpoint regions of two exemplary MFRs. The event on the left (right) exhibit coronal dimmings before (after) the flare onset, which is termed pre- and post-eruption dimming, respectively. Conjugated coronal dimmings (contours) that appear near flare ribbons are identified and tracked in the SDO/AIA 304~{\AA} passpand. (a1 \& b1) show composite images of AIA 304~{\AA} (red) and 1600~{\AA} (cyan) passbands. (a2 \& b2) show SDO/HMI vector magnetograms remapped with a cylindrical equal area projection. The background show $B_z$ within a range of [-1000, 1000] G, and the arrows show the transverse component of the magnetic field, with red (blue) arrows originating from positive (negative) polarity.  (a3 \& b3) show maps of $J_z$, the vertical component of electric current densities, as derived from HMI vector magnetograms. The $J_z$ maps are scaled within a range of [-40, 40] mA~m$^{-2}$ and averaged over two
    hours to minimize the noise. } 
\end{figure}

The footpoints of MFRs are identified by a pair of conjugate coronal dimmings that are observed in the vicinity of flare ribbons (Figure~\ref{fig:example}(a1, b1)) and associated with opposite polarities of photospheric magnetic field (Figure~\ref{fig:example}(a2, b2)). Coronal dimming regions may exhibit complex morphology, and care must be taken to discriminate the core and secondary dimmings: it is generally accepted that the former maps the footpoints of the eruptive structure, yet the latter maps the overlying field \cite[]{Dissauer2018}. Here coronal dimmings are identified and tracked in the 304~{\AA} passpand of the Atmospheric Imaging Assembly \citep[AIA;][]{Lemen2012} onboard SDO, and then dimmed pixels that persist for a long time are selected (marked by contours in Fig.~\ref{fig:example}, see Paper I for details), which are considered as the `core' footpoints of the MFR. The footpoint regions are then projected onto the vector magnetograms obtained by SDO/HMI and remapped with a cylindrical equal area projection, to measure the magnetic flux and electric currents within the regions. These dimmings may appear either before (9 events in Paper I, one of which is shown in the left column of Figure~\ref{fig:example}) or after the onset of flares (19 events in Paper I, one of which is shown in the right column of Figure~\ref{fig:example}), therefore being termed pre- or post-eruption dimming, respectively. 

An interesting result of Paper I is that 8 out of 28 MFRs carry significant non-neutralized currents, half of which are associated with pre-eruption dimming. The degree of neutralization, $R_z$, is measured by the magnitude ratio of direct current over return current, i.e.,
\[ R_z = \left|\frac{I_z^\mathrm{DC}}{I_z^\mathrm{RC}}\right|, \] 
where $I_z^\mathrm{DC}$ and $I_z^\mathrm{RC}$ are obtained by integrating separately the opposite signed values of $j_z=(\nabla\times\mathbf{B})_z/\mu_0$ over the identified footpoints of an MFR, with $I_z^\mathrm{DC}$ taking the dominant sign of $j_z/B_z$. This result prompted the authors to make a tentative conclusion that MFRs carrying substantial net current tend to form prior to eruptions. However, because of the small sample size, especially for the sample of MFRs with pre-erupion dimming, it is unclear how significant is the statistics and how strong is the degree of association. 

In this paper, we expand the sample size by further surveying all the two-ribbon flares of GOES class C5.0 and above occurring within 45$^\circ$ from the disk center from 2016 to the end of 2023. As a result, we find 9 more events (Table~\ref{tab:new}) that conform to our criteria in Paper I for MFR eruptions, with their footpoints clearly identified by conjugate coronal dimmings. With this expanded sample, we study the association between the MFRs with pre-eruption dimming and the non-neutralized electric current by non-parametric statistical tests. As an effort to dig out more information from the data, we further carry out a series of non-parametric tests on the associations among various physical variables, with emphasis on two groups, events with pre-eruption dimming versus those with post-eruption dimming, while Paper I focuses on events with non-neutralized electric current. The physical variables under investigation include the net magnetic flux $\Phi_\mathrm{net}$, $z$-component of net electric current $I_z^\mathrm{net}$, direct current $I_z^\mathrm{DC}$, and return current $I_z^\mathrm{RC}$, and degree of current neutralization $R_z$ (see Paper I for more details). The associations among these physical variables, however, have not been explored nor rigorously tested in Paper I.   

\begin{deluxetable*}{C|C|C|C|C|C|C|C|C}[!b]
	\tabletypesize{\scriptsize}
	\tablecaption{Magnetic properties at the footpoints of 9 MFRs from 2016 to 2023 \label{tab:new}}
	\tablecolumns{9}
	\tablewidth{0pt}
	\tablehead{
		\colhead{} & \colhead{} & \colhead{} & \colhead{} & \colhead{} & \colhead{} & \colhead{} & \colhead{} & \colhead{}  \\
		  \colhead{Date} & \colhead{Event} & \colhead{D-Type} & \colhead{FP Type} & \colhead{$\Phi_{net}$} & \colhead{$I_{z}^{net}$} & \colhead{$I_{z}^{DC}$} & \colhead{$I_{z}^{RC}$} & \colhead{$R_{z}$} \\
		\colhead{} & \colhead{} & \colhead{} & \colhead{} & \colhead{($10^{20}$ Mx)} & \colhead{($10^{11}$ A)} & \colhead{($10^{11}$ A)} & \colhead{($10^{11}$ A)} &  \colhead{} 
	}
	\startdata
	  2016-02-11 & C8.9 & Post & T & 5.68 & -0.22 & -5.94 & 5.72 & 1.0 \\
	               &      &      & L & -6.39 & 0.44 & 6.64 & -6.20 & 1.1 \\
	\hline
    2017-07-14 & M2.4 & Post & L & 4.68\tablenotemark{*} & -3.32 & -3.76 & 0.44 & \textbf{8.6} \\
	               &      &      & T & -26.38 & 6.16 & 29.90 & -23.74 & 1.3 \\
	\hline
    2017-09-04 & M1.7 & Pre & T & 35.65 & -13.12 & -25.86 & 12.73 & \textbf{2.0} \\
	               &      &     & L & -26.71 & 9.76 & 24.33 & -14.57 & 1.7 \\
	\hline
    2020-12-07 & C7.4 & Post & T & 10.80 & 0.12 & 9.06 & -8.94 & 1.0 \\
	               &      &      & L & -4.98 & -0.18 & -3.76 & 3.58 & 1.1 \\
	\hline
    2021-12-20 & X1.0 & Post & T & 21.04 & 0.41 & 21.67 & -21.26 & 1.0 \\
	               &      &      & L & -0.41 & 0.02 & 0.65 & -0.62 & 1.1 \\
	\hline
    2023-02-24 & M3.7 & Post & L & 2.73 & 0.20 & 7.37 & -7.17 & 1.0 \\
                  &      &      & T & -24.25 & -4.04 & -12.32 & 8.28 & 1.5 \\
    \hline
    2023-03-20 & FE & Post   & L & 36.31 & -3.13 & -24.52 & 21.39 & 1.2 \\
	               &    &        & T & -40.05 & -0.38 & -26.98 & 26.60 & 1.0 \\
	\hline
    2023-04-21 & M1.8& Post & T & 14.35 & -3.34 & -18.39 & 15.05 & 1.2 \\
                  &     &      & L & -33.59 & 2.52 & 30.77 & -28.25 & 1.1\\
        \hline
    2023-10-12 & C9.7 & Post & T & 28.06 & -7.35 & -25.86 & 18.51 & 1.4 \\
	               &      &      & L & -25.75 & -4.94 & -18.85 & 13.91 & 1.4 \\
	\enddata
    
	\tablecomments{`Pre' and `Post' in the 'D-type' column represent pre-eruption and post-eruption dimming, respectively. In the `Event' column, the GOES class of flares are given, except for a filament eruption. `L' and `T' in the `FP type' column represent the leading and trailing footpoint, respectively. Nonneutralized currents ($R_z\ge 2$) are boldfaced. Magnetic properties of MFRs from 2020 to 2015 are given in Table 2 (for pre-eruption dimming) and Table 3 (for post-eruption dimming) in Paper I.}
    \tablenotetext{*}{For the 2017-07-14 event, the small net flux for the foopoint of positive polarity is attributed to the dimming region being partly blocked by post flare loops.}
\end{deluxetable*}

\section{Statistical tests} \label{sec:Results}
The main objects of the statistical tests are as follows: 1) to test whether there exist significant differences between physical quantities measured at the conjugated footpoints of the same MFRs (\S\ref{subsec:conjugate}); 2) to test whether there exist significant differences between pre- and post-eruption dimming events (\S\ref{subsec:pre_vs_post}); 3) to test whether there exists a tendency for MFRs with pre-eruption dimming to be associated with non-neutralized currents (\S\ref{subsec:dimming}); 4) to test whether the different physical quantities are correlated with each other (\S\ref{subsec:correlation}), and 5) to test whether the three different approach to estimate the magnetic twist of MFRs give similar results (\S\ref{subsec:twist}). For item (1) Paper I has demonstrated that physical quantities at one footpoint is correlated with their counterparts at the other with various correlation coefficients, which we intend to put under more rigorous tests. Items (2, 4, 5) have not been explored in Paper I. Item (3) is a suggestion made in Paper I but has not been tested.

In the following statistical hypothesis testing, our strategy to adopt nonparametric methods is because of the small sample size and based on a series of tests demonstrating that the data does not approximately follow the normal distribution (Appendix~\ref{append:normality}). We follow the conventional practice by setting the significance level $\alpha$ to 5\% \cite[]{Cowles&Davis1982}, meaning that the null hypothesis, even though it is true, could still be rejected 5 times out of 100 samples. In other words, we believe in the null hypothesis on a 95\% confidence level. Meanwhile we obtain from the sample data the $p$-value \cite[]{Wasserstein&Lazar2016} that gives the probability of observing a result at least as extreme as the one obtained, given that the null hypothesis is true. One concludes that a result is statistically significant when $p < \alpha$, i.e., it reflects the characteristics of the whole population rather than the sampling error. Hence, a smaller $p$-value indicates that the data is more incompatible with the null hypothesis. On the other hand, a statistically insignificant result may also be important in physics. For example, the famous Michelson-Morley experiment found no significant difference between the speed of light in perpendicular directions \cite[]{Michelson&Morley1887}. 

\subsection{Hypothesis test on the differences between conjugated footpoints} \label{subsec:conjugate}
Naturally we expect that physical quantities measured at one footpoint are similar as those at the other footpoint of the same MFR (null hypothesis), but it is certainly not a straightforward impression that one may get by examining the values of each individual MFRs: they can be quite different. For example, the net magnetic flux measured at the trailing footpoint of the 2023-02-24 event is ten times larger than that at the leading one (Table~\ref{tab:new}). How significant are the differences?

We perform the Wilcoxon signed ranks test, which is a nonparametric statistical procedure for comparing two samples that are paired or related. Note that we rank the absolute value of the differences between paired data. $\sum R_+$ denotes the sum of the ranks with positive differences and $\sum R_-$ is the sum of the ranks with negative differences. The Wilcoxon $T$ is the smaller of $\sum R_+$ and $\sum R_-$.

With the level of significance setting at $\alpha=0.05$, the critical two-tailed value of $T$ for 10 pre-eruption and 27 post-eruption dimming events are 8 and 107, respectively. The null hypothesis can be rejected if the critical value equals or exceeds the obtained value. From Table~\ref{tab:wilcoxon}, one can see that for most physical quantities, our expectation stands true, except for the direct current of post-dimming events. 
 
\begin{deluxetable}{lCC} \label{tab:wilcoxon}
	\tablecaption{Wilcoxon signed ranks $T$ scores for the differences between conjugated footpoints}
	\tablecolumns{4}	
	\tablehead{\colhead{} & \colhead{Pre-eruption dimming} & \colhead{Post-eruption dimming} }
	
	\startdata  
	$\Phi_\mathrm{net}$ & 22 & 139  \\
	$I_z^\mathrm{net}$ & 27 & 152   \\
	$I_z^\mathrm{DC}$  & 25 & \textbf{105}   \\
	$I_z^\mathrm{RC}$  & 27 & 166   \\
	$R_z$  & 23 &  170  
	\enddata
	\tablecomments{At the level of significance $\alpha=0.05$, the critical value $T_c=8$ for the 10 events with pre-eruption dimming, $T_c$ = 107 for 27 events with post-eruption dimming. The $T$-score below the critical value is boldfaced.}	
\end{deluxetable}

\subsection{Hypothesis test on the differences between pre- and post-eruption dimming events} \label{subsec:pre_vs_post}
It is unclear whether dimming occurs before or during an eruption may be related to MFRs with different properties. We state the null hypothesis as follows: there is no tendency for physical quantities of pre-eruption dimming events to be significantly different than those of the post-eruption dimming events.

Assuming that the two samples are unrelated or independent, we perform the Mann-Whitney U-test. When combining and rank ordering the two samples together, the values from the two samples may be either randomly mixed or clustered at opposite ends. The former would mean that the two samples are not different, while the latter would indicate a difference between them. This is determined by two $U$ statistics as follows,
\begin{equation}
U_i=n_1n_2+\frac{n_i(n_i+1)}{2}-\sum R_i, \quad i=1,\ 2
\end{equation}
where $n_1=10$ and $n_2=27$ are the number of the pre- and post-eruption dimming events, respectively, and $\sum R_i$ is the sum of the ranks for each sample of interest. The smaller of the two $U$ statistics is then taken to calculate the $z$-score using a normal approximation, 
\[z =  \frac{U-\Bar{x}_U}{S_U},\]
with the mean, $\Bar{x}_U=n_1n_2/2$, and the standard deviation, $S_U=\sqrt{n_1n_2(n_1+n_2+1)/12}$. For a two-tailed test (i.e., we only care about whether there exists significant differences between the two groups) with $\alpha = 0.05$, we must not reject the null hypothesis if $-1.96 < z < 1.96$. 

The Mann Whitney U-test is applied to the unsigned average of various physical quantities measured at the two footpoints. From Table~\ref{tab:mann-whitney} one can see that, although the sum of the ranks for the MFRs with pre-eruption dimming is consistently smaller than that for the MFRs with post-eruption dimming, we cannot reject the null hypothesis for the majority of occasions, i.e., the obtained physical quantities of the pre-eruption dimming events are statistically similar to those of the post-eruption dimming events. The only occasion with $z$-score falling out of the range [-1.96, 1.96] is the direct current. We further consider the effect size to determine the degree of association between the two groups, i.e.,
\[ \rm ES = \frac{|z|}{\sqrt{n}},\]
where $n=37$ is the sample size. \cite{Cohen1988} defined the conventions for effect size as small = 0.10, medium = 0.30, and large = 0.50. The results shown in the last column of Table~\ref{tab:mann-whitney} indicate that on a medium level of association, $I_z^\mathrm{DC}$ for MFRs with pre-eruption dimming is smaller than those with post-eruption dimming.

\begin{deluxetable}{l cc cc cc} \label{tab:mann-whitney}
	\tablecaption{Mann-Whitney $U$-tests for the differences between MFRs with pre- and post-eruption dimming}
	\tablecolumns{8}	
	\tablehead{\colhead{} & \multicolumn{2}{c}{Pre-eruption} & \multicolumn{2}{c}{Post-eruption} & \multirow{2}*{$z$-score} & \multirow{2}*{ES} \\
    \colhead{} & \colhead{$R_1$} & \colhead{$U_1$} & \colhead{$R_2$} & \colhead{$U_2$} & \colhead{} & \colhead{} }
	
	\startdata  
	$\Phi_\mathrm{net}$	 & 225 & 100 & 478 & 179 & -1.20 & \\ 
	$I_z^\mathrm{net}$   & 238 & 87 & 465 & 183 & -1.64 &  \\	
	$I_z^\mathrm{DC}$    & 227 & 68 & 368 & 172 & \textbf{-1.97} & 0.32  \\ 
	$I_z^\mathrm{RC}$    & 216 & 109 & 487 & 161 & -0.89 &  \\	
	$R_z$                & 328 & 123 & 753 & 262 & -1.79 &  \\	 
	\enddata
	\tablecomments{The $z$-score that gives $p<\alpha$ is boldfaced.}	
\end{deluxetable}

\subsection{Hypothesis test on the tendency of pre-eruption dimming events to be non-neutralized} \label{subsec:dimming}
In Paper I, 8 of 28 events are identified to be associated with significant non-neutralized electric currents ($R_z\ge 2$ for either of the footpoints), half of which belong to pre-eruption dimming events. Among the newly found 9 events (Table~\ref{tab:new}), both 2017-07-14 and 2017-09-04 events have one footpoint with $R_z\ge2.0$. We now have 10 events in total that carry non-neutralized electric current, half of which are again associated with pre-eruption dimming. Here we intend to rigorously test the statistical significance of the observed tendency for pre-eruption dimming events to be non-neutralized. The test involves the statistical association between two categorical attributes, i.e., pre/post-eruption dimming versus degree of electric current neutralization, which is a typical occasion for the application of chi-square and Fisher exact tests. 

We assume that there are no differences between pre- and post-eruption dimming events on the degree of current neutralization (null hypothesis). The data are given in a $2\times 2$ contingency table (Table~\ref{tab:contingincy}). Two independent groups, pre- and post-eruption dimming events, were measured on the degree of current neutralization and classified as follows, i.e., neutralized if $R_z<2.0$ at both footpoints, and non-neutralized if $R_z\ge2.0$ at either footpoint. 

\begin{deluxetable}{lccc} \label{tab:contingincy}
	\tablecaption{Contingency Table for testing the tendency of current neutralization}
	\tablecolumns{4}	
	\tablehead{\colhead{} & \colhead{Pre-Eruption Dimming} & \colhead{Post-Eruption Dimming} & \colhead{Subtotal} }
	
	\startdata  
	Neutralized & 5 & 22 & 27 \\
	Non-Neutralized & 5 & 5 & 10 \\
	Subtotal & 10 & 27 & 37 
	\enddata	
\end{deluxetable}

\subsubsection{Chi-square test}
Pearson's chi-square statistic is performed as follows
\begin{equation}
\chi^2=\sum_j\sum_k\frac{(f_{ojk}-f_{ejk})^2}{f_{ejk}}
\end{equation}
where $f_{ojk}$ is the observed frequency for cell $A_jB_k$ and $f_{ejk}$ is the expected frequency for the same cell, which is obtained by dividing the product of the row total and the column total by the grand total, $N$. The degrees of freedom, df, for the chi-square are 
\[ \mathrm{df} = ( R - 1) ( C - 1 ) = 1, \]
where $R$ is the number of rows and $C$ is the number of columns. We obtained $\chi^2=3.67$; the corresponding $p$-value is 0.055, which slightly exceeds the level of significance set at $\alpha=0.05$. Technically we cannot reject the null hypothesis.

\subsubsection{Fisher exact test}
One caveat for the chi-square test is that the sample for pre-eruption dimming events is small. For such situations, the Fisher exact test is often recommend. The $p$-value of the Fisher exact statistic is given by
\begin{equation}
p=\frac{(A+B)!(C+D)!(A+C)!(B+D)!}{N!\,A!\,B!\,C!\,D!}=0.058,
\end{equation}
where $A=5$, $B=22$, $C=5$, $D=5$, and $N=37$ is the total number of events, as given in Table~\ref{tab:contingincy}.

%Meanwhile, we construct a $2\times2$ contingency table for hypothetical data that represent a more extreme occurrence than that obtained, i.e., $A=0$, $B=19$, $C=9$, $D=0$. The corresponding Fish exact statistic is
%\[ p_2= 1.448\times10^{-7}\approx0\]
%The probability is found by adding the two results above.
%\[p=p_1+p_2=0.16\]
The obtained $p$-value is not small enough to reject the null hypothesis, which is consistent with the chi-square test, suggesting that the difference between the pre- and post-eruption dimming events about the degree of current neutralization is not statistically significant at the 5\% level of significance, but are significant at 10\% level. However, these tests do not describe the strength of the association. We again adopt the effect size as a measure of association between the nominal variables of the $2\times 2$ contingency table. The effect size, ranging from 0 to 1, can be represented with the $\phi$ coefficient \cite[]{Cohen1988}, i.e.,
\[\phi=\sqrt{\frac{\chi^2}{N}} = 0.31.\]
This suggests that there is a medium level of association between the pre-eruption dimming events and the non-neutralized current at the footpoints.

\subsection{Spearman rank-order correlation between different magnetic properties of MFRs} \label{subsec:correlation}
For two paired variables on an ordinal scale of measurement with a sample size $n\geq4$, their relationship can be measured by the Spearman rank-order correlation. When there exist no ties in the ranked values, we obtain the Spearman rank-order correlation coefficient, $r_s$, as follows, 
\begin{equation}
r_s=1-\frac{6\sum_{i=1}^{n}D_i^2}{n(n^2-1)},
\end{equation}
where $D_i$ is the difference between the ranks of the two paired variables. The critical value for a sample size of $n=37$ and a level of significance at $\alpha=0.05$ is 0.325, i.e., if the critical value is less than or equal to the obtained $|r_s|$, then we can reject the null hypothesis that there is no significant correlation between two physical quantities measured at the footpoints of MFRs. 

Here we take the unsigned mean values measured at the footpoints to perform the Spearman rank-order correlation. The results as given in Table~\ref{tab:spearman} suggest that there is a significant correlation between the net flux $\Phi_\mathrm{net}$ and the electric current parameters, including $I_z^\mathrm{net}$, $I_z^\mathrm{DC}$, and $I_z^\mathrm{RC}$. We suggest that the correlation between magnetic flux and electric currents is most likely because strong electric currents are usually detected in regions of strong magnetic field (e.g., Fig.~\ref{fig:example}), with the limited resolution and sensitivity of contemporary magnetographs. That the strong currents tend to be clustered may also explain the strong correlation between $I_z^\mathrm{DC}$ and $I_z^\mathrm{RC}$. $R_z$ is positively correlated with the net current $I_z^\mathrm{net}$ and negatively correlated with the return current $I_z^\mathrm{RC}$, which may be a natural result, since $R_z$ is the ratio of direct over return current and $I_z^\mathrm{net}$ is the subtraction between direct and return current. On the other hand, it is not easy to understand the lacking of significant correlations between $I_z^\mathrm{net}$ and $I_z^\mathrm{RC}$, between $I_z^\mathrm{DC}$ and $R_z$, and between $\Phi_\mathrm{net}$ and $R_z$, which might require further tests on a larger sample.

\begin{deluxetable}{lccccc} \label{tab:spearman}
	\tablecaption{Spearman rank-order correlation coefficient}
	\tablecolumns{6}	
	\tablehead{\colhead{} & \colhead{$\Phi_\mathrm{net}$} & \colhead{$I_z^\mathrm{net}$} & \colhead{$I_z^\mathrm{DC}$} & \colhead{$I_z^\mathrm{RC}$} & \colhead{$R_z$}   }
	
	\startdata  
	$\Phi_\mathrm{net}$ & & \textbf{0.63} &  \textbf{0.89} & \textbf{0.74} & 0.17 \\
	$I_z^\mathrm{net}$  & &      &  \textbf{0.48} & 0.13 & \textbf{0.79}  \\
	$I_z^\mathrm{DC}$   & & & &  \textbf{0.89} & -0.06 \\
	$I_z^\mathrm{RC}$   & & & &                 & \textbf{-0.41} \\	
	\enddata
	\tablecomments{The critical value is 0.325 for a significance level of 0.05 and a sample size of 37. Correlation coefficients that bear statistical significance are boldfaced.}	
\end{deluxetable}

\subsection{Hypothesis test on the different approaches to estimate the magnetic twist} \label{subsec:twist}
Two alternative definitions of magnetic twist are given in \citet[][their Eqs.~13 and 16]{Berger&Prior2006} and their relation is clarified in \citet[][Appendix C]{Liu2016}: the former is precise, depending on geometrical quantities; the latter is approximate, depending on physical quantities. What is often used in practice is the latter, i.e., 
\begin{equation} \label{eq:tw_orig}
T_w=\int \frac{\mu_0J_\parallel}{4\pi B}\,d\ell,
\end{equation} which is integrated along each individual magnetic field line.

Based on Eq.~\ref{eq:tw_orig}, Paper I employed three different approaches to estimate the average magnetic twist of MFRs. Namely,
\begin{subequations} \label{eq:tw}
	\begin{eqnarray}
	T_{w,1} =& \left\langle\cfrac{B_\theta}{2\pi rB_z}\right\rangle L \\
	T_{w,2} =& \left\langle\cfrac{\mu_0J_z}{4\pi B}\right\rangle L \\
	T_{w,3} =& \cfrac{\mu_0 I}{4\pi\Phi}L
	\end{eqnarray}
\end{subequations}
where $L$ is given by the separation of the MFR's footpoints, assuming a half-circular shape, $r$ is the distance to the geometric center of each individual footpoint, $I$ is the  electric current flowing through the footpoint, and $\Phi$ is the magnetic flux through the footpoint. In the first two formulas, the physical quantities inside the angle brackets are avearged over all the pixels inside the identified footpoint regions. In the third, $T_{w,3}$ can be given either by the net current or the direct current, denoted as $T_{w,3}^\mathrm{net}$ and $T_{w,3}^\mathrm{DC}$, respectively. Thus, 4 different values of magnetic twist are given for each of the 10 MFRs that carry substantial non-neutralized currents (Table~\ref{tab:twist});  8 MFRs before 2016 are reproduced from Table 4 in Paper I. Here, we will test if any one of these approaches is different from others. 

\begin{deluxetable*}{CCCCCCCCCCC}[!b]
	\tabletypesize{\scriptsize}
	\tablecaption{Magnetic twist of MFRs with Non-neutralized Current \label{tab:twist}}
	\tablecolumns{10}
	\tablewidth{0pt}
	\tablehead{
		  \colhead{Date} & \colhead{Distance} & \colhead{Length} & \colhead{D-type} & \colhead{FP sign} & \colhead{FP type} & \colhead{$T_{w,1}$} & \colhead{$T_{w,2}$} & \colhead{$T_{w,3}^\textrm{net}$} & \colhead{$T_{w,3}^\textrm{DC}$}
		}
	\startdata
    2011-08-02 & 64.8 & 101.7 & Post & + & T & 1.8$\pm{0.6}$ & 1.5$\pm{0.6}$ & 0.8$\pm{0.1}$ & 1.2$\pm{0.0}$ \\
	  &      &       &      & - & L & 2.6$\pm{0.6}$ & 2.0$\pm{0.7}$ & 0.6$\pm{0.2}$ & 1.6$\pm{0.1}$ \\
	\hline
	2011-09-30 & 26.7 & 42.0 & Pre & + & T & 1.8$\pm{0.6}$ & 1.9$\pm{0.7}$ & 0.3$\pm{0.1}$ & 0.9$\pm{0.1}$ \\
	  &      &      &     & - & L & 2.0$\pm{0.7}$ & 0.8$\pm{0.5}$ & 0.2$\pm{0.0}$ & 0.3$\pm{0.0}$ \\
	\hline
	2012-06-14 & 96.8 & 152.1 & Pre & + & L & 2.4$\pm{0.6}$ & 0.7$\pm{0.3}$ & 0.4$\pm{0.1}$ & 0.6$\pm{0.0}$ \\
	  &      &       &     & - & T & 1.6$\pm{0.6}$ & 1.7$\pm{0.7}$ & 1.0$\pm{0.0}$ & 1.9$\pm{0.0}$ \\
	\hline
	2013-08-12 & 53.0 & 83.3 & Post & + & L & 2.3$\pm{0.6}$ & 1.8$\pm{0.7}$ & 0.4$\pm{0.2}$ & 1.0$\pm{0.1}$ \\
	  &      &      &      & - & T & 1.6$\pm{0.6}$ & 0.9$\pm{0.6}$ & 0.2$\pm{0.1}$ & 0.6$\pm{0.1}$ \\
	\hline
	2013-10-13 & 47.9 & 75.3 & Post & + & L & 1.8$\pm{0.6}$ & 1.7$\pm{0.7}$ & 0.4$\pm{0.3}$ & 1.0$\pm{0.1}$ \\
	  &      &      &      & - & T & 1.4$\pm{0.6}$ & 1.6$\pm{0.6}$ & 0.7$\pm{0.1}$ & 1.5$\pm{0.1}$ \\
	\hline
	2014-08-25 & 58.3 & 91.5 & Pre & + & T & 1.7$\pm{0.6}$ & 1.9$\pm{0.7}$ & 0.8$\pm{0.2}$ & 1.7$\pm{0.1}$ \\
	  &      &      &     & - & L & 2.2$\pm{0.5}$ & 1.5$\pm{0.6}$ & 0.5$\pm{0.2}$ & 1.1$\pm{0.0}$ \\
	\hline
	2014-09-08 & 64.2 & 100.9 & Pre & + & L & 1.8$\pm{0.5}$ & 1.0$\pm{0.6}$ & 0.4$\pm{0.0}$ & 0.5$\pm{0.0}$ \\
	  &      &       &     & - & T & 1.1$\pm{0.4}$ & 1.7$\pm{0.7}$ & 0.1$\pm{0.0}$ & 0.8$\pm{0.0}$ \\
	\hline
	2015-06-22 & 97.1 & 152.5 & Post & + & T & 1.8$\pm{0.6}$ & 1.8$\pm{0.7}$ & 0.7$\pm{0.1}$ & 1.5$\pm{0.0}$ \\
	&      &       &      & - & L & 1.4$\pm{0.6}$ & 2.0$\pm{0.7}$ & 0.4$\pm{0.2}$ & 1.8$\pm{0.1}$  \\
	\hline
	2017-07-14 & 126.3 & 198.3 & Post & + & L & 4.7$\pm{0.3}$ & 0.9$\pm{0.8}$ & 1.4$\pm{0.3}$ & 1.6$\pm{0.1}$\\
	&     &     &      & - & T & 1.3$\pm{0.5}$ & 2.2$\pm{0.8}$ & 0.5$\pm{0.1}$ & 2.3$\pm{0.1}$ \\
	\hline
	2017-09-04 & 83.2 & 130.7 & Pre & + & T & 1.8$\pm{0.7}$ & 1.2$\pm{0.6}$ & 0.5$\pm{0.0}$ & 1.0$\pm{0.1}$ \\
	&      &       &     & - & L & 2.0$\pm{0.7}$ & 1.7$\pm{0.7}$ & 0.5$\pm{0.0}$ & 1.0$\pm{0.1}$ \\
	\enddata
	\tablecomments{ For $T_{w,1}$ and $T_{w,2}$, the uncertainties are estimated from the standard deviation of the twist per unit length $\tau$, which is calculated for every pixel in FP+ and FP-. For $T_{w,3}$, the uncertainties are given through error propagation. }
\end{deluxetable*}

For this purpose, we perform the Friedman test, which is used to compare more than two dependent/related samples. First, we rank the $T_w$ values for each event. If there are no ties in the ranks, we compute the Friedman test statistic, $Fr$, as follows,
\begin{equation}
Fr = \frac{12}{nk(k+1)}\sum_{i=1}^k R_i^2 - 3n(k+1),
\end{equation}
where $k=4$ is the number of approaches, $n=10$ is the number of MFRs, and $R_i$ is the sum of the ranks from the $i$th approach. However, if there are any ties in the ranks we have to determine $Fr$ as follows
\begin{equation}
Fr=\cfrac{n(k-1)(\sum_{i=1}^k\frac{R_i^2}{n} - C_F)}{\sum_i\sum_j r_{ij}^2 - C_F},
\end{equation}
where $C_F=\dfrac14nk(k+1)^2$ is the ties correction and $r_{ij}$ is the rank corresponding to event $i$ and approach $j$. 

\begin{deluxetable}{lccccc} \label{tab:friedman-wilcoxon}
	\tablecaption{Hypothesis tests on different approaches to estimate magnetic twist $T_w$}
	\tablecolumns{6}	
	\tablehead{  & \multicolumn{2}{c}{Friedman test} & \multicolumn{3}{c}{Wilcoxon test\tablenotemark{c}}\\
		\colhead{FP} & \colhead{Four approaches\tablenotemark{a}}  & \colhead{Three approaches\tablenotemark{b}} & \colhead{$T_{w,1}$ vs $T_{w,2}$}  & \colhead{$T_{w,1}$ vs $T_{w,3}^\mathrm{DC}$} &\colhead{$T_{w,2}$ vs $T_{w,3}^\mathrm{DC}$} }
	\startdata  
	+ & \textbf{23.42} & \textbf{13.00} & \textbf{7} & \textbf{0}  & \textbf{7}\\
	- & \textbf{19.56} & 2.60 & 23 & 15 &  \textbf{4}\\
	m &\textbf{23.88} & \textbf{9.80} & \textbf{5} & \textbf{1} & \textbf{6}
	\enddata
	\tablenotetext{a}{$Fr$ values for the four approaches, $T_{w,1}$, $T_{w,2}$, $T_{w,3}^\mathrm{net}$, and $T_{w,3}^\mathrm{DC}$. The critical $Fr$ value for $k=4$ and $n=10$ is 7.80 at the significance level of 0.05.}
	\tablenotetext{b}{$T_{w,3}^\mathrm{net}$ is excluded. The critical $Fr$ value for $k=3$ and $n=10$ is 6.20 at the significance level of 0.05.}
	\tablenotetext{c}{The critical $T$-score is 8 for a two-tailed test, with the sample size $n=10$ and a level of significance $\alpha=0.05$.}
	\tablecomments{$Fr$ values above the critical value and $T$-scores below the critical value, which will lead us to reject the null hypothesis, are shown in boldface. Data are taken from Table~\ref{tab:twist}.}	
\end{deluxetable}

We obtained $Fr$ values for $T_w$ estimated from footpoints with positive and negative polarities separately, as well as for the average $T_w$ from the conjugated footpoints. We found that hey are all above the critical $Fr$ value for $k=4$ and $n=10$, which is 7.80 at the significance level of 0.05 (Table~\ref{tab:friedman-wilcoxon}). This indicates that we must reject the null hypothesis that the four different approaches give similar results. 

In Paper I, it has been noted that $T_{w,3}^\mathrm{net}$ tend to give smaller values than the other three approaches. We then re-performed the Friedman test by excluding this method. The resultant $Fr$ values, except for the footpoints of negative polarity, are still above the critical value for $k=3$ and $n=10$, which is 6.20 at the significance level of 0.05 (Table~\ref{tab:friedman-wilcoxon}). These values support the rejection of the null hypothesis that the three $T_w$ formulas (Eq.~\ref{eq:tw}) give similar results. 

Further, we performed Wilcoxon signed-ranks tests (see also \S\ref{subsec:conjugate}) to compare any two of the three approaches (Table~\ref{tab:friedman-wilcoxon}). The results suggest that except for the negative-polarity footpoints they are all different from each other, with $T$-scores below the critical value of 8 for a sample of 10 events and a significance level of 0.05. It can be concluded that overall the three different approaches do give different results.

\section{Conclusion \& Discussion} \label{sec:C&D}
At the significance level of 5\%, we can make the following conclusions, in terms of magnetic properties measured at the footpoints of MFRs: 
\begin{itemize}
    \item There exists no significant differences between conjugated footpoints of the same MFR. This shows that our method of identifying the MFR footpoints through coronal dimming gives reasonable results. Had we found statistically significant differences between the MFR footpoints, it would have cast doubt on the reliability of this method.
    \item There exist no significant differences between MFRs with pre-eruption dimming and those with post-eruption dimming, except that on a medium level of association, the direct current for MFRs with pre-eruption dimming is smaller than those with post-eruption dimming. More importantly, there exists a medium level of association between pre-eruption dimming and non-neutralized electric current.  
    \item There exist significant correlations between net magnetic flux and electric current parameters including net current, DC and RC. This may result from an empirical fact that strong electric currents are usually detected in regions of strong magnetic fields.
    \item Using different formulas (Eq.~\ref{eq:tw}) to estimate the magnetic twist of MFRs with physical quantities measured at footpoints yield significantly different results, which suggests that an MFR is not a monolithic structure and it is important to take into account the spatially inhomogeneous distribution of electric current density and magnetic field to determine the magnetic twist of MFRs.
\end{itemize}

\subsection{MFRs with pre-eruption vs post-eruption dimming}
Our statistical test corroborates Paper I's suggestion that MFRs with pre-eruption dimming are more likely to be associated with non-neutralized electric current than those that produce only post-eruption dimming. Since in the pre-eruption dimming events, the dimming typically lasts for hours and is associated with the gradual expansion of a coronal structure and the same structure erupts subsequently \citep[e.g.,][]{WangW2019}, there is little doubt that the MFR has already formed prior to the eruption. We further surmise that an MFR carrying substantial net current is subject to strong Lorentz-self force, also known as hoop force \cite[]{Kliem&Torok2006}, which points radially outward for a bent current channel; such an MFR is hence more likely to expand and rise gradually. The density decrease in an expanding MFR is able to produce coronal dimming that persists for hours before eruption \cite[e.g.,][]{WangW2019}. On the other hand, it is possible for an MFR to form prior to the eruption but remain stationary due to strong confining force exerted by the overlying field. In this case, the MFR would only produce post-eruption dimming. This may explain why there is no significant differences between the magnetic properties of MFRs with pre-eruption dimming and those with post-eruption dimming.

For an MFR embedding a filament, mass draining may have an effect on coronal dimming. During the slow-rise phase, mass drainage toward the footpoints of the filament might compensate for the density decrease due to the MFR expansion and make the dimming less visible. During the fast-rise phase, mass drainage often produces brightening in EUV at the impact site \cite[e.g.,][]{Gilbert2013}, where coronal dimming could even be reversed. It may be worth investigating how mass draining and coronal dimming compete with each other in filament eruptions.

\subsection{Magnetic twist of MFRs}
In estimating the magnetic twist of MFRs from magnetic properties at their footpoints, we did not distinguish the electric current perpendicular from that parallel to the magnetic field, but assume that $I_z$ is the current flowing into the higher atmosphere. Alternatively, magnetic twist of MFRs can be estimated from extrapolated nonlinear force-free fields \cite[e.g.,][]{Liu2016}, in which electric currents are approximately field-aligned. Supposedly, the magnetic field and electric current would adjust themselves and evolve from a non-force-free state in the photosphere to a force-free state in the corona. But it is unclear how the photospheric current would relate to the coronal current in an MFR and which approach to obtain the magnetic twist gives a better approximation. These issues require further investigation from both observational and modeling perspective.

\acknowledgements We are grateful to the anonymous referee whose constructive comments and suggestions help improve the manuscript. This work was supported by the NSFC (11925302, 42188101, 12373064, and 42274204) and the National Key R\&D Program of China (2022YFF0503002).

%\bibliography{statistic}{}
%\bibliographystyle{aasjournal}

\clearpage
\appendix
\section{Hypothesis tests on the normality of data} \label{append:normality}
In scientific practice, one usually employs parametric statistical tests, e.g., the Student's $t$-test; however, such tests are based on the assumption that data samples are adequately large and randomly drawn from a population approximately resembling a normal distribution. Here, because of the small sample size, it is not recommended to take the parametric approach for statistical tests involving MFRs with pre-eruption dimming (10 events) or MFRs carrying substantial electric currents (10 events). Below, we further test whether the normality assumption is valid for our data, employing kurtosis and skewness tests as well as the Kolmogorov-Smirnov test; if not, we must turn to nonparametric tests \cite[]{nonparametric_book}, which do not rely on the assumption of normal distribution.  

\subsection{Testing kurtosis and skewness of the sample normality}
First we use the kurtosis and skewness to test if our sample resembles a normal distribution. 
The kurtosis measure how flat or peaked a sample is relative to a normal distribution. A positive kurtosis indicates that the sample is concentrated near the center of the distribution, while a negative kurtosis indicates that the sample is relatively flat. The so-called excess kurtosis, $K$, and standard error of kurtosis, SE$_K$, are found by
\begin{subequations}
\begin{eqnarray}
K=\left[\frac{n(n+1)}{(n-1)(n-2)(n-3)}\sum_{i=1}^n\left(\frac{x_i-\bar{x}}{s}\right)^4\right] - \frac{3(n-1)^2}{(n-2)(n-3)} \\
\mathrm{SE}_K=\sqrt{\frac{24n(n-1)^2}{(n-2)(n-3)(n+5)(n+3)}}
\end{eqnarray}	
\end{subequations}
where $\bar{x}$ is the sample mean and $s$ is the sample standard deviation. The $z$-score for the kurtosis is
\[z_K=\frac{K-0}{\mathrm{SE}_K}\]

The skewness measures the horizontal symmetry of a sample with respect to a normal distribution. A sample distribution has a negative (positive) skewness, if it is concentrated to the right (left) side of the normal distribution. The skewness, $S$, and standard error of the skewness, SE$_S$, are found by 
\begin{subequations}
	\begin{eqnarray}
	S=\frac{n}{(n-1)(n-2)}\sum_{i=1}^n\left(\frac{x_i-\bar{x}}{s}\right)^3 \\
	\mathrm{SE}_S=\sqrt{\frac{6n(n-1)}{(n-2)(n+1)(n+3)}}
	\end{eqnarray}	
\end{subequations}
The $z$-score for the skewness is
\[z_S=\frac{S-0}{\mathrm{SE}_S}\]

Comparing the calculated $z$-scores to the values of the normal distribution, we get the $p$-values of the tests on various physical quantities measured at the footpoints of MFRs (Table~\ref{tab:normality}). For the desired significance level of $\alpha = 0.05$, we can reject the null hypothesis that the sample has an approximately normal distribution, if  $p < 0.05$. From Table~\ref{tab:normality}, one can see that although we cannot reject the null hypothesis for the small sample of pre-eruption dimming events, it is clear that the distribution of post-eruption dimming events and of the data as a whole are statistically different from the normal distribution. 

\subsection{Kolmogorov-Smirnov test}
The Kolmogorov-Smirnov one-sample test is a procedure to examine the agreement
between two sets of values. Here we examine the normality of the collected samples by comparing the observed frequency distribution against the empirical frequency distribution which is based on a normal distribution. Using the point at which these two cumulative frequency distributions show the largest divergence, the Kolmogorov-Smirnov test gives the $p$-value of a two-tailed probability estimate, which helps to determine if our sample resembles a normal distribution.

The Kolmogorov-Smirnov test statistic, $Z$, is given by
\begin{equation}
Z=\sqrt{n}D_\mathrm{max},
\end{equation}
where $n$ is the number of values in the observed sample and $D_\mathrm{max}$ is the largest absolute value divergence between the observed and empirical cumulative frequency distributions.
Then, the $p$-value is determined as follows:
\begin{subequations}
\begin{eqnarray}
p=& 1, &\quad \mathrm{if}\quad 0\leq Z <0.27 \\
p=& 1-\frac{2.506628}{Z}(Q+Q^9+Q^{25}), &\quad \mathrm{if}\quad 0.27 \leq Z<1 \\
p=& 2(T-T^4+T^9-T^{16}), &\quad \mathrm{if}\quad 1 \leq Z<3.1 \\
p=& 0, &\quad \mathrm{if}\quad Z \geq 3.1
\end{eqnarray}
\end{subequations}
where 
\[Q=\exp(1.233701 Z^{-2})\]
and
\[T=\exp(-2Z^2)\]
The null hypothesis is that the observed sample has an approximately normal distribution.  We can reject the null hypothesis if the obtained $p$-value is smaller than the level of significance setting at $\alpha=0.05$. Similarly to the kurtosis and skewness tests, one can see from Table~\ref{tab:normality} that although we cannot reject the null hypothesis for the small sample of pre-eruption dimming events, it is clear that the distribution of post-eruption dimming events and of the data as a whole are statistically different from the normal distribution. Hence, we must employ non-parametric statistical methods.

\begin{deluxetable}{lc ccc ccc ccc} \label{tab:normality}
	\tablecaption{$p$-values obtained from tests on the normality of data}
	\tablecolumns{11}	
	\tablehead{\colhead{} & \multirow{2}*{FP} & \multicolumn{3}{c}{Pre-eruption dimming} & \multicolumn{3}{c}{Post-eruption dimming} & \multicolumn{3}{c}{All} \\
		\colhead{} & \colhead{} & \colhead{Kurtosis} & \colhead{Skewness} & \colhead{KS} &
		\colhead{Kurtosis} & \colhead{Skewness} & \colhead{KS} & \colhead{Kurtosis} & \colhead{Skewness} & \colhead{KS} }
	
	\startdata  
	\multirow{2}*{$\Phi_\mathrm{net}$} & + & 0.69 & 0.13 & 0.59 & 0.00 & 0.00 & 0.00 & 0.04 & 0.00 & 0.00 \\
						  & - & 0.60 & 0.14 & 0.08 & 0.87 & 0.05 & 0.00 & 0.86 & 0.04 & 0.00 \\						  
	\multirow{2}*{$I_z^\mathrm{net}$} & + & 0.71 & 0.08 & 0.63 & 0.00 & 0.00 & 0.04 & 0.00 & 0.00 & 0.00\\
									  & - & 0.92 & 0.47 & 0.10 & 0.32 & 0.00 & 0.00 & 0.71 & 0.01 & 0.00 \\
	\multirow{2}*{$I_z^\mathrm{DC}$}  & + & 0.61 & 0.25 & 0.00 & 0.00 & 0.00 & 0.01 & 0.00 & 0.00 & 0.00\\
									  & - & 0.68 & 0.34 & 0.00 & 0.27 & 0.02 & 0.00 & 0.55 & 0.04 & 0.00 \\
	\multirow{2}*{$I_z^\mathrm{RC}$}  & + & 0.02 & 0.00 & 0.50 & 0.00 & 0.00 & 0.00 & 0.00 & 0.00 & 0.00\\
									  & - & 0.86 & 0.44 & 0.01 & 0.03 & 0.00 & 0.00 & 0.20 & 0.01 & 0.00 \\
	\multirow{2}*{$R_z$}  & + & 0.25 & 0.04 & 0.46 & 0.00 & 0.00 & 0.00 & 0.00 & 0.00 & 0.00 \\
						  & - & 0.17 & 0.01 & 0.30 & 0.00 & 0.00 & 0.00 & 0.00 & 0.00 & 0.00 
	\enddata
%	\tablenotetext{a}{Kolmogorov-Smirnov test}
	\tablecomments{The `KS' columns give the results of the Kolmogorov-Smirnov test. The column `FP' indicates the polarity sign. }	
\end{deluxetable}

\end{document}